\documentclass[letterpaper]{JHEP3}

\usepackage{amsmath,amsthm, amssymb}

\newcommand{\dst}{\displaystyle}

\newcommand{\be}{\begin{equation}}
\newcommand{\ee}{\end{equation}}
\newcommand{\ba}{\begin{array}}
\newcommand{\ea}{\end{array}}

\newcommand{\bea}{\begin{eqnarray}}
\newcommand{\eea}{\end{eqnarray}}
\newcommand{\bma}{\begin{matrix}}
\newcommand{\ema}{\end{matrix}}
\newcommand{\nn}{\nonumber}
\newcommand{\mc}{\mathcal}
\newcommand{\beq}{\stackrel{+0}{=}}

\newcommand{\st}{\sqrt2}
\newcommand{\cd}{\mathcal D}
\newcommand{\p}{\partial}
\newcommand{\Uij}{U_i{}^j}

\newcommand{\ov}{\overline}
\newcommand{\wh}{\widehat}
\newcommand{\wt}{\widetilde}
\newcommand{\Eta}{\mathcal H}
\newcommand{\psibar}{\ov \psi}
\newcommand{\etabar}{\ov \eta}
\newcommand{\sibar}{\ov \sigma}
\newcommand{\labar}{\ov \la}
\newcommand{\xibar}{\ov \xi}
\newcommand{\tabar}{\ov \ta}

\newcommand{\ep}{\varepsilon}
\newcommand{\al}{\alpha}
\newcommand{\la}{\lambda}
\newcommand{\da}{\delta}
\newcommand{\Ga}{\Gamma}
\newcommand{\La}{\Lambda}
\newcommand{\Si}{\Sigma}
\newcommand{\si}{\sigma}
\newcommand{\ta}{\theta}

\title{\bf Boundary conditions in \\ the Mirabelli and Peskin model}
\author{Dmitry V. Belyaev\footnote{Present address:
DESY-T, Notkestrasse 85, 22603 Hamburg, Germany}
\\
Department of Physics and Astronomy, The Johns Hopkins University, \\
3400 North Charles Street, Baltimore, MD 21218, USA\\ 
E-mail: \email{dmitry.belyaev@desy.de}}

\preprint{}

\abstract{
We show how the (globally supersymmetric) model of Mirabelli
and Peskin can be formulated in the boundary 
(``downstairs'' or ``interval'') picture. 
The necessary Gibbons-Hawking-like terms appear naturally
when using (codimension one) superfields.
This formulation is free of the $\delta(0)$ ambiguities 
of the orbifold (``upstairs'') picture while
describing the same physics since the boundary conditions
on the fundamental domain are the same. 
The (natural) boundary conditions follow from the variational
principle and form a closed orbit under supersymmetry variation.
They reduce to the ``odd $=0$'' boundary conditions
in the absence of bulk-boundary coupling.
We emphasize that
the action is supersymmetric \emph{without} the use of any
boundary conditions in the \emph{off-shell} formulation
(but some boundary conditions are necessary for
on-shell supersymmetry!).
}

\begin{document}
\section{Introduction}
In the last decade there was a revival of interest in theories
on a manifold with boundary. It started from a ground-breaking
paper by Horava and Witten \cite{hw}, who showed that 
eleven-dimensional supergravity on a manifold with boundary
appears as one of the low energy limits of the heterotic
string theory. Soon after, Mirabelli and Peskin \cite{mp}
introduced a simplified construct of a five-dimensional 
globally supersymmetric Yang-Mills theory coupled to a
four-dimensional hypermultiplet living on the boundary. 
They argued that the analysis
of this simpler system can provide useful insights into 
various aspects of the more complicated Horava-Witten theory.

Another fruitful research direction started with the work
of Randall and Sundrum \cite{rs1, rs2}. They showed that
if our four-dimensional world is localized on a three-brane
in a five-dimensional space-time with negative cosmological
constant, then there is an attractive geometrical solution to the
weak/Planck hierarchy problem.
This scenario was later supersymmetrized
\cite{abn, gp1, flp1}. In Ref.~\cite{bb1} it was shown how
to unify the original approaches to the supersymmetric
Randall-Sundrum scenario, and the issue of supersymmetric
boundary conditions was addressed.

The original idea for the work presented here
and in the companion paper, Ref.~\cite{my2},
was to understand why the
use of boundary conditions was essential in the supersymmetric
Randall-Sundrum scenario, but apparently was not
important in the Mirabelli and Peskin model. 
The results of this research show that both statements require
some adjustments. In this paper we will show that the use of
some boundary conditions is required for supersymmetry of
the Mirabelli and Peskin model, while in Ref.~\cite{my2} we find
that supersymmetry of the Randall-Sundrum scenario requires
the use of \emph{fewer} boundary conditions than
was previously assumed.

Another motivation for this work 
was to understand the structure of boundary conditions
in the Mirabelli and Peskin model, in particular,
their closure under supersymmetry.
We present here a set of Neumann-like boundary conditions
(which we call ``natural'' following Ref.~\cite{barth}),
which guarantee that
the general variation of the action vanishes for \emph{arbitrary}
variations of the fields on the boundary.
(That is the boundary conditions are derived exactly as
the bulk equations of motion.)
These natural boundary conditions reduce to ``odd~$=0$''
ones (the vanishing of the odd fields on the boundary)
\emph{only} when there is no coupling of the bulk fields
to the boundary. We emphasize that it is inconsistent to
assume the ``odd~$=0$'' boundary conditions when the coupling 
is present (unless it is only a zeroth-order approximation in
the perturbative calculations).

As was explained by Horava and Witten \cite{hw}, any theory
on a manifold with boundary can be equivalently described as
a theory on an orbifold, when a mirror image of the fundamental
domain is introduced with the corresponding identifications.
This leads to two descriptions of the same theory:
the ``downstairs'' (boundary) picture and 
the ``upstairs'' (orbifold) picture.
In the boundary picture we speak of ``boundary,'' in the
orbifold picture we have a ``brane'' (a set of fixed points
of the mirror reflection).
Both pictures have some advantages, but the boundary picture
is more fundamental.

The Mirabelli and Peskin model \cite{mp} was originally set up in
the orbifold picture. One of the results of Ref.~\cite{mp}
was to provide an algorithm for coupling bulk fields to 
brane-localized fields in a supersymmetric way.
In this paper, we show how the model can be formulated in
the boundary picture. We eliminate the mirror image space
and work directly on the fundamental domain with boundary.
The boundary conditions follow, as in the orbifold picture,
from the variational principle provided we include
a special Gibbons-Hawking-like
boundary term~\cite{gibh} (we call it ``$Y$-term'' to acknowledge 
the work of York \cite{york1,york2}).
This boundary term is also required by supersymmetry.
We show how the $Y$-term
follows naturally when using the codimension one ($D=4$, $N=1$) 
superfields, which makes them very useful in constructing
supersymmetric bulk-boundary couplings.
Eliminating auxiliary fields, one then obtains on-shell formulation
which is free of $\da(0)$ ambiguities present in the orbifold picture. 

The paper is organized as follows.
In Section \ref{sec-mp1}, we review the original formulation 
of the Mirabelli and Peskin model: off-shell, in components,
on the orbifold. In Section \ref{sec-mp2}, we derive the
natural boundary conditions and discuss their closure under
supersymmetry. In Section \ref{sec-mp3}, we turn to the 
superfield formulation of the model and show how the boundary
conditions are reproduced. In Section \ref{sec-mp4}, we
present the formulation of the model in the boundary picture.
We discuss there the derivation and the role of the $Y$-term.
In Section~\ref{sec-mp5}, we discuss the on-shell case
(when the auxiliary fields of the model are eliminated).
There we show that in the orbifold picture, as in
the off-shell case, we still do not need to use any boundary conditions
to establish supersymmetry, whereas in the boundary picture,
the use of some boundary conditions is required.
We also show that the ``odd~$=0$'' boundary conditions are consistent
only as \emph{free field} boundary conditions.  
Finally, in Section \ref{sec-mp6},
we discuss the transition between the boundary and the orbifold
pictures. 

The Appendices contain some technical details, which are
separated for clarity of discussion. The basic conventions are
the same as in Ref.~\cite{bb1}.

\section{The component formulation}
\label{sec-mp1}
In this section we write the model of Mirabelli and Peskin,
adjusting it to our conventions. We present the model in its
original formulation: on the orbifold, in components, with
off-shell field content. In the later sections we present
the model in other settings: on a space-time domain with boundary,
in the superfield formulation, and truncated to 
on-shell field content.
For simplicity, we consider only the Abelian case, since it is
sufficient for our purposes.

\subsection{Bulk Lagrangian}
The bulk Lagrangian is the standard 
globally supersymmetric Lagrangian
for an Abelian gauge multiplet, $(A_M, \Phi, \La_i, X_a)$,
in five dimensions
($M,N=\{0,1,2,3,5\}$, $i,j=\{1,2\}$, $a=\{1,2,3\}$),
\bea
\label{L5}
\mc{L}_5=-\frac{1}{4}F_{MN}F^{MN}-\frac{1}{2}\p_M\Phi\p^M\Phi
-\frac{i}{2}\wt\La^i\Ga^M\p_M\La_i+\frac{1}{2}X_a X_a \; .
\eea
Here $F_{MN}$ is the field strength for a gauge boson $A_M$,
$\Phi$ is a scalar, $\La_i$ is a symplectic-Majorana spinor,
and $X_a$ is a triplet of auxiliary fields. 
The Lagrangian has four global symmetries
(supersymmetry, translation, Lorentz and $SU(2)$),
and one local invariance ($U(1)$ gauge invariance).
The Lorentz transformation is standard and it is not important
for our discussion. The other transformations are
\begin{itemize}
\item
(global) supersymmetry (with fermionic parameter $\Eta_i=\rm{const}$),
\bea
\label{MPsusytr}
\ba{lcl}
\da_\Eta A^M &=& i\wt\Eta^i\Ga^M\La_i \\[5pt]
\da_\Eta\Phi &=& i\wt\Eta^i\La_i \\[5pt]
\da_\Eta X_a &=& \wt\Eta^i(\si_a)_i{}^j\Ga^M\p_M\La_j\\[5pt]
\da_\Eta\La_i &=& (\Si^{MN}F_{MN}+\Ga^M\p_M\Phi)\Eta_i
+i X_a(\si_a)_i{}^j\Eta_j \; ;
\ea
\eea
\item
(global) translation (with parameter $v^M=\rm{const}$),
\bea
\da_v (A_M, \Phi, X_a, \La_i)=v^K\p_K(A_M, \Phi, X_a, \La_i) ;
\eea
\item
(global) $SU(2)$ rotation (with constant matrix parameter $U\in SU(2)$),
\bea
\La_i^\prime = \Uij\La_j, \qquad
X_a{}^\prime \si_a=U(X_a \si_a)U^\dagger ;
\eea
\item
(local) $U(1)$ gauge transformation (with parameter $u(x)$),
\bea
\da_u A_M=\p_M u \; .
\eea
\end{itemize}
(In the above, the $\si_a=\{\si_1,\si_2,\si_3\}$ 
are the usual Pauli matrices.)
The Lagrangian $\mc{L}_5$ is invariant ($\da\mc{L}_5=0$) under
the $U(1)$, $SU(2)$ and Lorentz transformations, but it changes
into a total derivative under the translation and
supersymmetry transformations,
\bea
\da_v \mc{L}_5 = \p_M (v^M \mc{L}_5), \qquad
\da_\Eta\mc{L}_5 = \p_M \wt K^M,
\eea
where $\wt K^M$ is given in Eq.~(\ref{wtKM}). 
The supersymmetry algebra has the following form,
\bea
[\da_\Xi, \da_\Eta] =v^K\p_K +\da_u \; ,
\eea
where the parameters of the translation and the $U(1)$ transformation
are
\bea
v^K = 2i(\wt\Eta^i\Ga^K\Xi_i), \qquad
u = -2i(\wt\Eta^i\Ga^K\Xi_i)A_K-2i(\wt\Eta^i\Xi_i)\Phi \; .
\eea

\subsection{Breaking $N=2$ to $N=1$ supersymmetry}
We will use the two-component spinor notation, in which the 
symplectic-Majorana spinors $\La_i$ and $\Eta_j$ are represented
by pairs of two-component spinors: $(\la_1, \la_2)$
and $(\eta_1, \eta_2)$ (see Appendix~\ref{app-comp}).
After the $M=\{m,5\}$ split
and in the two-component spinor notation, the Lagrangian
assumes the following form,
\bea
\mc{L}_5 &=& -\frac{1}{4}F_{mn}F^{mn}-\frac{1}{2}F_{m5}F^{m5}
-\frac{1}{2}\p_m\Phi\p^m\Phi -\frac{1}{2}\p_5\Phi\p_5\Phi
+\frac{1}{2}X_{12}X_{12}^\ast +\frac{1}{2}X_3^2 \nn\\
&&-\left[
\frac{i}{2}\la_1\si^m\p_m\labar_1
+\frac{i}{2}\la_2\si^m\p_m\labar_2
+\frac{1}{2}(\la_2\p_5\la_1-\la_1\p_5\la_2) +h.c.
\right],
\eea
where $X_{12}=X_1+i X_2$.
The Lagrangian is invariant 
(up to total $\p_m$ and $\p_5$ derivatives)
under the supersymmetry transformations with arbitrary (constant)
$\eta_1$ and $\eta_2$.
The Lagrangian is, therefore, $N=2$ supersymmetric.

In the presence of a brane/boundary, we can keep only half of the
bulk supersymmetry intact. This statement is based on the following
standard argument. The commutator of two supersymmetry transformations
generates a translation with parameter
\bea
v^M =2i(\wt\Eta^i\Ga^K\Xi_i)\; .
\eea
The brane breaks translational invariance, allowing only $v^5=0$.
(We assume that the brane/boundary is located at $x^5=\rm{const}$.)
For the supersymmetry parameters this implies
\bea
v^5=2(\eta_2\xi_1-\eta_1\xi_2)+h.c.=0 \; .
\eea
Therefore, the two supersymmetry parameters have to be related,
\bea
\eta_2=\al\eta_1 \; , 
\eea
where $\al$ is an arbitrary complex constant. 
This eliminates one linear combination of $\eta_1$ and $\eta_2$;
the orthogonal linear combination describes the unbroken
$N=1$ supersymmetry.

Note that such
supersymmetry transformations still generate a non-zero $U(1)$ 
transformation with
\bea
u=-2[i(\eta_1\si^m\xibar_1+\eta_2\si^m\xibar_2)+h.c.]A_m \; .
\eea
This implies that the brane/boundary action we will introduce must
be gauge invariant (for the total action to be supersymmetric).

We choose to preserve $\eta_1$ supersymmetry,
so that from now on we set 
\bea
\eta_2=0 \; .
\eea
Any other choice of $\al$ (preserving another linear combination 
of $\eta_1$ and $\eta_2$) can be obtained by a global $SU(2)$ rotation
(see Appendix~\ref{app-su2}).

Under the $N=1$ ($\eta=\eta_1$) supersymmetry, the five-dimensional
gauge supermultiplet splits into two four-dimensional 
supermultiplets (see Appendix~\ref{app-n1}), 
the gauge and chiral multiplets,
\bea
(v_m, \la, D) &=& (A_m, \la_1, X_3-\p_5\Phi)\\[5pt]
(\phi_2, \psi_2, F_2) &=& (\Phi+i A_5, -i\st\la_2, -X_{12}) \; .
\eea
The bulk Lagrangian can also be written in terms of these
fields,
\bea
\label{L5VF}
\mc{L}_5 &=&
-\frac{1}{4}v_{mn}v^{mn}
-\frac{1}{2}(\p_5 v_m)(\p_5 v^m)
-\frac{1}{2}\p_m\phi_2\p^m\phi_2^\ast
+\frac{1}{2}F_2F_2^\ast
+\frac{1}{2}D^2
\nn\\
&&-\frac{i}{2}(\p_m\phi_2-\p_m\phi_2^\ast)\p_5 v^m
+\frac{1}{2}D\p_5(\phi_2+\phi_2^\ast)
\nn\\
&&-\left(
\frac{i}{2}\la\si^m\p_m\labar
+\frac{i}{4}\psi_2\si^m\p_m\psibar_2
+\frac{i}{2\st}(\psi_2\p_5\la-\la\p_5\psi_2)+h.c.
\right).
\eea

\subsection{Brane Lagrangian}
In the orbifold picture, the bulk action is invariant 
under supersymmetry because the
total derivative terms integrate to zero. Therefore, to have a
supersymmetric bulk-plus-brane system we simply need to add a
supersymmetric brane action, containing couplings between the 
induced bulk fields and intrinsic brane fields.

We consider a four-dimensional chiral supermultiplet $(\phi, \psi, F)$
living on the brane and couple it to the bulk vector supermultiplet
$(v_m, \la, D)$. The brane Lagrangian is therefore the standard
four-dimensional Lagrangian \cite{wb}, coupling the two multiplets 
in a gauge invariant way,\footnote{\label{footQ}
The coupling constant $Q$ can be made explicit by multiplying
every $v_m$, $\la$ and $D$ in $\mc{L}_4$ by $Q$. 
The right-hand sides of the boundary conditions, Eqs.~(\ref{bc1})
and (\ref{bc2}), are then also multiplied by $Q$, so that turning
the coupling off (setting $Q=0$) gives the ``odd $=0$'' boundary conditions.
In the non-abelian case \cite{mp}, the strength of the coupling
is dependent on the coupling constant $g$ of self-interaction
of the gauge fields; the decoupling happens when the brane chiral
multiplets are in the trivial representation $R$ of the gauge
group: $t^A_{ab}=0$.}
\bea
\label{L4}
\mc{L}_4=
-\cd_m\phi\cd^m\phi^\ast
-i\psibar\sibar^m\cd_m\psi
+F F^\ast
+\frac{i}{\st}(\phi^\ast\la\psi-\phi\labar\psibar)
+\frac{1}{2}\phi\phi^\ast D \; ,
\eea
where $\cd_m=\p_m+\frac{i}{2}v_m$ .

The brane Lagrangian is invariant (up to a total $\p_m$
derivative) under the standard supersymmetry transformations,
see Eqs.~(\ref{strV}) and (\ref{strF}). It is also invariant
under the $U(1)$ gauge transformation (see Appendix~\ref{app-n1}),
\bea
\da_u v_m=\p_m u, \qquad
\da_u (\phi, \psi, F)=-\frac{i}{2} u (\phi, \psi, F) \; .
\eea

\section{Boundary conditions}
\label{sec-mp2}
We will consider first the orbifold picture, where the bulk
action is $\mc{L}_5$ integrated over $\mathbb{R}^{1,4}$ and
the brane action is $\mc{L}_4$ integrated over the hypersurface
$x^5=0$. In this picture, both total $\p_m$ and $\p_5$ derivatives 
can be neglected. The bulk and brane actions, therefore, are
\emph{separately} supersymmetric, which is true without reference
to any jump/boundary conditions. However, the presence of the
brane-localized sources (due to the brane action) requires
certain jump conditions across the brane to be satisfied.
And when we impose the $\mathbb{Z}_2$ symmetry, these jump
conditions turn into boundary conditions on each side of the
brane. In this section we derive these boundary conditions
and discuss their closure under the $N=1$ 
supersymmetry transformations.\footnote{
The closure under supersymmetry of boundary conditions 
in various supersymmetric theories
was discussed before. Some of the early references are
Refs.~\cite{dive1,igarashi}. For a more recent discussion
see Refs.~\cite{global,nieu}.}

\subsection{Primary boundary conditions}
In the orbifold picture, the bulk and brane Lagrangians can be
combined into a total bulk-plus-brane Lagrangian,
\bea
\mc{L}=\mc{L}_5+\da(z)\mc{L}_4 \; ,
\eea
where the delta-function $\da(z)$ is localized at $z\equiv x^5=0$.
The equations of motion for this Lagrangian are straightforward to
derive and are summarized in Eq.~(\ref{EOM}). 
The $\da(z)$-terms in the equations enforce the 
following jump conditions,
\bea
\ba{lcl}
[F_{m5}] &=&
-\dst \frac{i}{2}(\phi\cd_m\phi^\ast-\phi^\ast\cd_m\phi)
-\frac{1}{2}\psi\si_m\psibar\\[8pt]
{[} \Phi {]} &=&
\dst -\frac{1}{2}\phi\phi^\ast\\[8pt]
{[} \la_2 {]} &=&
\dst -\frac{i}{\st}\phi^\ast\psi \; ,
\ea
\eea
where the square brackets denote the jump across the brane,
\bea
[\Phi(x)]\equiv\Phi(x,z=+0)-\Phi(x,z=-0) \; .
\eea

Let us now introduce a $\mathbb{Z}_2$ parity, $f(-z)=P[f]f(+z)$,
according to
\bea
\label{parity}
P[A_m, \la_1, X_3]=+1, \quad
P[A_5, \la_2, X_{12}, \Phi]=-1 \; .
\eea
(As usual, we call $P[f]=+1$ fields ``even,''
and $P[f]=-1$ fields ``odd''.)
These parity assignments are consistent with the equations of motion
and the supersymmetry transformations. They allow us to rewrite
the jump conditions as boundary conditions for fields at $z=+0$,
\bea
\label{bc1}
\ba{lclcl}
B_1(A_m) &:& 2F_{m5} &\beq&
-\dst \frac{i}{2}(\phi\cd_m\phi^\ast-\phi^\ast\cd_m\phi)
-\frac{1}{2}\psi\si_m\psibar 
\\[8pt]
B_1(\Phi) $$ &:& $$ 2\Phi &\beq&
\dst -\frac{1}{2}\phi\phi^\ast 
\\[8pt]
B_1(\la_1) $$ &:& $$ 2\la_2 &\beq&
\dst -\frac{i}{\st}\phi^\ast\psi \; .
\ea
\eea

The bulk-plus-brane equations of motion split into bulk equations of
motion (for the bulk fields), brane equations of motion (for the brane
fields, both intrinsic and induced from the bulk) 
and boundary conditions (relating near-brane values of the bulk
fields to the brane fields). For a general variation of the 
bulk-plus-brane action to vanish, all the equations of motion and the
boundary conditions must be satisfied. On the other hand, the variation
of the bulk-plus-brane action under the supersymmetry transformations
vanishes \emph{without} the use of either equations of motion or
boundary conditions.

\subsection{Secondary boundary conditions}
If supersymmetry is a true symmetry of the bulk-plus-brane system,
the boundary conditions should also be invariant under 
the supersymmetry transformations.
The primary boundary conditions $B_1(A_m)$, $B_1(\Phi)$ and $B_1(\la_1)$
do not form a supersymmetric system of equations. However, after
a finite number of the supersymmetry variations, 
we arrive at a supersymmetric
system including both the primary and secondary boundary conditions.

The structure of the supersymmetry variation for this system of boundary
conditions is as follows,
\bea
\label{structure}
\ba{lclclcl}
\da_\eta B_1(\Phi) &=& B_1(\la_1) && && \\[3pt]
\da_\eta B_1(\la_1) &=& B_1(A_m) &\oplus&  B_2(\la_1)
&\oplus&  B_1(\Phi)\\[3pt]
\da_\eta B_2(\la_1) &=& B_2(A_m) &\oplus& B_1(\la_1) &&  \\[3pt]
\da_\eta B_1(A_m) &=& B_2(A_m) &\oplus&  B_1(\la_1) &&\\[3pt]
\da_\eta B_2(A_m) &=& B_3(A_m) &\oplus&  B_1(A_m) && \\[3pt]
\da_\eta B_3(A_m) &=& B_2(A_m) \; . && && 
\ea
\eea
The secondary boundary conditions are
\bea
\label{bc2}
\ba{lclcl}
B_2(A_m) &:& 2\p_5\la_1 &\beq&
\dst \frac{1}{\st}\si^m\psibar\cd_m\phi+\frac{i}{\st}\psi F^\ast
+\frac{1}{2}\phi\phi^\ast\la_1 \\[10pt]
B_2(\la_1) &:& 2X_{12} &\beq& F\phi^\ast \\[8pt]
B_3(A_m) &:& 2\p_5 D &\beq& \mc{L}_4^r \; ,
\ea
\eea
where $\mc{L}_4^r$ is real and differs from $\mc{L}_4$
by a total $\p_m$ derivative,
\bea
\label{L4r}
\mc{L}_4^r &=&
-\cd_m\phi\cd^m\phi^\ast
+F F^\ast
+\frac{1}{2}\phi\phi^\ast D\nn\\
&&-\frac{i}{2}(\psi\si^m\cd_m\psibar-\cd_m\psi\si^m\psibar)
+\frac{i}{\st}(\phi^\ast\la\psi-\phi\labar\psibar) \; .
\eea
Note that the boundary conditions are gauge invariant.

The fact the the boundary conditions are closed under
supersymmetry implies that they can be cast in a superfield form.
This is indeed so, as will be shown in the next section.

\section{Superfield description}
\label{sec-mp3}
$N=1$ supersymmetry can be conveniently described in terms of
superfields \cite{wb}. 
For the Mirabelli and Peskin model, the superfield
description of the unbroken $N=1$ supersymmetry was discussed
in Ref.~\cite{agw}. 
In this section, we use this description to cast our 
boundary conditions in a superfield form. We also find
that the superfield description leads to a bulk
Lagrangian $\mc{L}_5{}^\prime$ different from $\mc{L}_5$ by
a total derivative term, which produces the necessary 
boundary Lagrangian in the boundary picture.

\subsection{Bulk and brane Lagrangians}
\label{superL}
The $N=2$ five-dimensional supersymmetric theory can be described
in terms of $N=1$ four-dimensional superfields. 
The vector supermultiplet $(v_m, \la, D)$ and chiral supermultiplets
$(\phi, \psi, F)$ and $(\phi_2, \psi_2, F_2)$ are described by a real
vector superfield $\bf V$ (in the WZ gauge) and chiral superfields
$\bf\Phi$ and $\bf\Phi_2$, respectively. The component expansions are
given in Appendix~\ref{app-sf}.
The gauge transformation, Eqs.~(\ref{gtr1}) and (\ref{gtr2}),
generalizes to a supergauge transformation, parametrized by a
chiral superfield $\bf\La$,
\bea
\label{sgauge}
\bf
\da V=\La+\La^\dagger, \quad
\da\Phi=-\La\Phi, \quad
\da\Phi_2=2\p_5\La \; .
\eea

Let us consider a supergauge invariant Lagrangian that can be
built from the superfields $\bf V$ and $\bf\Phi_2$,
\bea
\label{L5pr}
\mc{L}_5{}^\prime =
\frac{1}{4}\int d^2\ta \; {\bf W}{\bf W} + h.c.
+\int d^2\ta d^2\tabar \; {\bf Z}^2 \; ,
\eea
where $\bf W$ is the field strength for $\bf V$ (see Ref.~\cite{wb}), 
and
\bea
{\bf Z}=\p_5 {\bf V}-\frac{1}{2}({\bf\Phi_2}+{\bf\Phi_2}^\dagger)
\eea
is defined following Ref.~\cite{heb}. 
Both $\bf W$ and $\bf Z$ are invariant
under the supergauge transformation, Eq.~(\ref{sgauge}).
Expanding in components and comparing with Eq.~(\ref{L5VF}), we find
\bea
\mc{L}_5{}^\prime = \mc{L}_5
+\p_5\left(-\frac{1}{2}(\phi_2+\phi_2^\ast)D
-\frac{i}{2\st}(\la\psi_2-\labar\psibar_2)\right)
+\frac{1}{16}\p_m\p^m(\phi_2+\phi_2^\ast)^2 \; .
\eea
Or, in terms of the original bulk fields,
\bea
\label{p5inL5}
\mc{L}_5{}^\prime = \mc{L}_5
+\p_5\left(-\Phi D-\frac{1}{2}(\la_1\la_2+h.c.)\right)
+\frac{1}{4}\p_m\p^m(\Phi^2) \; .
\eea
In the orbifold picture both total derivatives can be neglected.
But, as we will see in the next section,
in the boundary picture the total $\p_5$ derivative gives 
rise to an important boundary term.

The brane action can also be written in the superfield form. 
The following Lagrangian,
\bea
\mc{L}_4{}^{\prime}=
\int d^2\ta d^2\tabar \; {\bf\Phi}^\dagger e^{\bf V}{\bf\Phi} \; ,
\eea
differs from $\mc{L}_4$, Eq.~(\ref{L4}), by a total $\p_m$ derivative,
\bea
\label{L4L4pr}
\mc{L}_4{}^{\prime}
&=& \mc{L}_4^r+\frac{1}{4}\p_m\p^m(\phi\phi^\ast) \nn\\
&=& \mc{L}_4
+\frac{1}{4}\p_m\p^m(\phi\phi^\ast)
-\frac{i}{2}\p_m(\psi\si^m\psibar) \; .
\eea
Therefore they both lead to the same brane action.

It is clear that 
using the $D=4, N=1$ superfields keeps the $N=1$ ($\eta=\eta_1$) 
supersymmetry manifest. 
A less obvious observation is that under the $N=1$ supersymmetry, 
the bulk Lagrangian $\mc{L}_5{}^\prime$ varies
into a total $\p_m$ (not $\p_M$) derivative term, that is
\bea
\fbox{\text{
$\da_\eta\mc{L}_5{}^\prime$ {\it does not} contain a $\p_5$ term.
}}
\nn
\eea
(This is so because for the $D=4, N=1$ superfields
$x^5$ is just a parameter. The highest component of a superfield
varies into a total derivative, which is $\p_m$
for our superfields.)
This ensures that $\mc{L}_5{}^\prime$ automatically gives
rise to a supersymmetric action both in the orbifold \emph{and}
boundary pictures!

\subsection{Boundary conditions in the superfield form}
In the orbifold picture, the (superfield) bulk-plus-brane Lagrangian is 
\bea
\mc{L}^\prime =
\mc{L}_5{}^\prime+\mc{L}_4{}^\prime\da(z) &=&
\frac{1}{4}\int d^2\ta \; {\bf W}{\bf W}+ h.c.
\nn\\
&&+\int d^2\ta d^2\tabar \left\{
{\bf Z}^2
+{\bf\Phi}^\dagger e^{\bf V}{\bf\Phi}\da(z)\right\},
\eea
where 
${\bf Z}=\p_5 {\bf V}-\frac{1}{2}({\bf\Phi_2}+{\bf\Phi_2}^\dagger)$.
One can derive the equations of motion directly from varying the
superfields \cite{wb}.
The variation of chiral superfields requires some care. 
But in our case all boundary conditions come from varying
the vector superfield $\bf V$. Keeping only terms with $\p_5$
and $\da(z)$, we obtain,
\bea
\da\mc{L}^\prime=\int d^2\ta d^2\tabar \Big\{
\da{\bf V}\Big[ -2\p_5{\bf Z}
+{\bf\Phi}^\dagger e^{\bf V}{\bf\Phi}\da(z) \Big]
+\p_5\Big[ 2{\bf Z}\da{\bf V} \Big] \Big\}+\dots
\eea
The total $\p_5$ derivative is irrelevant in the orbifold
picture. (But, as we will see, it is essential in the derivation 
of boundary conditions in the boundary picture.)
From the equation of motion,
\bea
2\p_5{\bf Z} =
{\bf\Phi}^\dagger e^{\bf V}{\bf\Phi}\da(z)+\dots \; ,
\eea
and assuming the parity assignments (\ref{parity})
(so that $\bf V$ is even, while $\bf\Phi_2$ and $\bf Z$ are odd),
we obtain the following boundary condition,
\bea
\label{bcsf}
2{\bf Z} \equiv
2\p_5{\bf V}-({\bf\Phi_2}+{\bf\Phi_2}^\dagger) \beq
\frac{1}{2}{\bf\Phi}^\dagger e^{\bf V}{\bf\Phi} \; .
\eea
Using the component expansions (see Appendix \ref{app-compexp}),
we can split this superfield boundary condition 
into the following relations for the component fields,
\bea
\ba{lcrcl}
1 &:& -(\phi_2+\phi_2^\ast) &\beq&
\dst \frac{1}{2}\phi\phi^\ast \\[8pt]
\ta &:& -\st\psi_2 &\beq&
\dst \frac{1}{\st}\phi^\ast\psi \\[8pt]
\ta^2 &:& -F_2 &\beq&
\dst \frac{1}{2}\phi^\ast F \\[8pt]
\ta\si^m\tabar &:& 
-2\p_5 v_m-i(\p_m\phi_2-\p_m\phi_2^\ast) &\beq&
\dst -\frac{i}{2}(\phi\cd_m\phi^\ast-\phi^\ast\cd_m\phi)
-\frac{1}{2}\psi\si_m\psibar \\[8pt]
\tabar{}^2\ta &:&
\dst -2i\p_5\la-\frac{i}{\st}\si^m\p_m\psibar_2 &\beq&
\dst -\frac{1}{2}\la\phi\phi^\ast
+\frac{1}{\st}\psi F^\ast\\ && &&
\dst +\frac{i}{2\st}\si^m(\phi\cd_m\psibar-\psibar\cd_m\phi)
\\[8pt]
\ta^2\tabar{}^2 &:&
\dst \p_5 D-\frac{1}{4}\p_m\p^m(\phi_2+\phi_2^\ast) &\beq&
\dst \frac{1}{2}\mc{L}_4^r+\frac{1}{8}\p_m\p^m(\phi\phi^\ast) \; .
\ea
\nn\\
\eea
It is easy to check that these boundary conditions are equivalent
to Eqs.~(\ref{bc1}) and (\ref{bc2}). 
See also Appendix~\ref{app-bcs}.

Note that in the superfield approach
\emph{all} boundary conditions appear as \emph{primary} boundary 
conditions and their supermultiplet structure is manifest.

The superfield derivation 
also explains how boundary conditions with and without $\p_5$
can appear in the same supermultiplet. When $\bf V$ is not in the
WZ gauge, $\p_5$ appears in every boundary condition,
but fixing the gauge allows one to eliminate terms with $\p_5$ 
acting on the pure gauge degrees of freedom.

\section{Boundary picture}
\label{sec-mp4}
In this section we will discuss supersymmetry and boundary
conditions as they appear in
the boundary picture, where our space-time domain is
$\mc{M}=\mathbb{R}^{1,3}\times[0,+\infty)$. 
The space-time now has a boundary $\p\mc{M}$ 
at $z=0$. We no longer have to deal with singularities,
but now the total $\p_5$ derivatives cannot be neglected.
With our setting, we have
\bea
\int_{\mc{M}}\p_M K^M =
\int_{\mc{M}}\p_5 K^5 =
\int_{\p\mc{M}}(-K^5) \; .
\eea
The measures of integration, $d^5x$ on $\mc{M}$ and
$d^4x$ on $\p\mc{M}$, are implicit.

\subsection{The action}
In the boundary picture, the bulk-plus-boundary action which leads
to the same superfield boundary condition (\ref{bcsf}) is
\bea
\label{bryS}
S=\int_{\mc{M}}\mc{L}_5{}^\prime
+\frac{1}{2}\int_{\p\mc{M}}\mc{L}_4{}^{\prime} \; .
\eea
(We can replace here $\mc{L}_4{}^{\prime}$ by $\mc{L}_4$ since
they differ only by a total $\p_m$ derivative, Eq.~(\ref{L4L4pr}),
which integrates to zero on $\p\mc{M}$.)
Indeed, a general variation of the vector superfield $\bf V$ gives
the following boundary term,
\bea
\da S=\int_{\p\mc{M}}\int d^2\ta d^2\tabar
\da{\bf V}\left(-2{\bf Z}
+\frac{1}{2}{\bf\Phi}^\dagger e^{\bf V}{\bf\Phi}\right).
\eea
Requiring this term to vanish for arbitrary $\da\bf V$, enforces
the boundary condition (\ref{bcsf}),
\bea
2{\bf Z}\beq\frac{1}{2}{\bf\Phi}^\dagger e^{\bf V}{\bf\Phi} \; .
\eea

\subsection{Important boundary term}
The two bulk Lagrangians, $\mc{L}_5$ and $\mc{L}_5{}^\prime$, 
defined in Eqs.~(\ref{L5}) and (\ref{L5pr}), respectively,
lead to two \emph{different} actions,
\be
S_5=\int_{\mc{M}}\mc{L}_5 \quad \text{and} \quad
S_5{}^\prime = \int_{\mc{M}}\mc{L}_5{}^\prime \; ,
\ee
because of the $\p_5$ term in Eq.~(\ref{p5inL5}).
The actions differ by a boundary term. Namely,
\bea
\label{L5+GH}
S_5{}^\prime = S_5
+\int_{\p\mc{M}}\left(\Phi D+\frac{1}{2}(\la_1\la_2+h.c.)\right) .
\eea
We know that $S_5{}^\prime$ is supersymmetric, because $\mc{L}_5{}^\prime$ 
was constructed out of ($D=4$, $N=1$) superfields (see the last
remark in Section \ref{superL}). On the other hand, the action
$S_5$, based on the original bulk Lagrangian (\ref{L5}), is \emph{not}
supersymmetric in the boundary picture. The extra boundary term
is \emph{required} to make the bulk action supersymmetric.

\subsection{Supersymmetry of the action}
It is instructive to check supersymmetry of the 
action $S_5{}^\prime$ explicitly. We have
\bea
\da_\eta S_5{}^\prime = \int_{\p\mc{M}}
\left(-\wt K^5+\da_\eta(\Phi D)
+\frac{1}{2}(\da_\eta(\la_1\la_2)+h.c.)\right) ,
\eea
where $\wt K^M$ is given in Eq.~(\ref{wtKM}). The boundary
term can be rewritten as
\bea
F^{5n}\da_\eta A_n +\Phi\da_\eta D+X_3\da_\eta\Phi
+(\la_2\da_\eta^{\prime\prime}\la_1+\la_1\da_\eta^{\prime}\la_2+h.c.) \; ,
\eea
where $\da_\eta^{\prime\prime}\la_1=i X_3\eta$ and
$\da_\eta^{\prime}\la_2=-(i F_{m5}+\p_m\Phi)\si^m\etabar$.
It is easy to check that this boundary term is a total $\p_m$ derivative.
This explicitly shows that the action $S_5{}^\prime$ 
is supersymmetric,
\bea
\da_\eta S_5{}^\prime = 0 \; .
\eea
We conclude that the total bulk-plus-boundary action (\ref{bryS}),
\bea
S=\int_{\mc{M}}\mc{L}_5+\int_{\p\mc{M}}Y
+\frac{1}{2}\int_{\p\mc{M}}\mc{L}_4 \; ,
\eea
where
\bea
\label{Yterm}
Y=\Phi D+\frac{1}{2}(\la_1\la_2+h.c.) \; ,
\eea
is $N=1$ supersymmetric. This statement {\it does not}
rely on using any boundary conditions. 
The ``improved'' bulk action $S_5{}^\prime$ (the sum of terms with
$\mc{L}_5$ and $Y$),
and the boundary action (with $\mc{L}_4$) 
are {\it separately} supersymmetric.

\subsection{Variational principle}
As we will now show, the boundary $Y$-term plays the role of the 
Gibbons-Hawking term for our bulk action.
(We chose letter $Y$ to honor York \cite{york1, york2}, 
whose name could as well be
included in the phrase ``Gibbons-Hawking term''.)

Using Eq.~(\ref{gvarL5}), we find that the general variation of the
original bulk action $S_5$ has the following
boundary term,
\bea
\da S_5 = \int_{\p\mc{M}}(-K^5)
=\int_{\p\mc{M}}\left(\da A_n F^{5n}+\da\Phi\p_5\Phi
+\frac{1}{2}(\la_2\da\la_1-\la_1\da\la_2+h.c.)\right).
\eea
This expression, however, gets modified when we make the 
following field redefinition,
\bea
X_3 \quad \longrightarrow \quad D=X_3-\p_5\Phi \; .
\eea
Considering $D$ (rather than $X_3$) as an independent
bulk field, we find
\bea
\da S_5=\int_{\p\mc{M}}\left(\da A_n F^{5n}-\da\Phi D
+\frac{1}{2}(\la_2\da\la_1-\la_1\da\la_2+h.c.)\right).
\eea
The analogous general variation for the ``improved'' 
action (\ref{L5+GH})
is
\bea
\da S_5{}^\prime=\int_{\p\mc{M}}\left(\da A_n F^{5n}+\Phi\da D
+(\la_2\da\la_1+h.c.)\right).
\eea
This expression contains variations of only those (combinations of the)
bulk fields which we include in the boundary action.
(Just as the gravitational action with the Gibbons-Hawking 
boundary term \cite{gibh} 
contains only variations of the metric field on the
boundary, but not variations of its normal derivative.)
Adding the contribution from the variation of the
boundary action $\int_{\p\mc{M}}\mc{L}_4$ and requiring the total 
expression to vanish for arbitrary 
variations $\da A_m$, $\da D$ and $\da\la_1$, we obtain the
primary boundary conditions (\ref{bc1}).\footnote{
For comparison, see 
the derivation and discussion of boundary conditions 
in the boundary (``interval'') picture in 
Refs.~\cite{csaki1, csaki2}.}

The boundary $Y$-term, therefore, plays two roles at the same time:
\begin{itemize}
\item[1)]
It makes the ``improved'' bulk action $S_5{}^\prime$ supersymmetric.
\item[2)]
It makes the variational principle well-defined 
(the equations of motion and the boundary conditions follow 
from the vanishing of the general variation
of the action for arbitrary field variations).
\end{itemize}

\section{On-shell case}
\label{sec-mp5}
This section is of particular importance for the discussion
of (on-shell) five-dimensional supergravity on a manifold
with boundary given in the companion paper, Ref.~\cite{my2}.

In this section we will show that after eliminating the auxiliary
fields only the total bulk-plus-brane action is supersymmetric
(and not the bulk and the brane actions separately, 
as is the case in the off-shell formulation). 
In addition, in the on-shell \emph{boundary}
picture, supersymmetry of the total action \emph{does} rely
on using (some of) the boundary conditions!

\subsection{On-shell in the orbifold picture}
The total bulk-plus-brane Lagrangian is
\be
\mc{L}=\mc{L}_5+\da(z)\mc{L}_4 \; ,
\ee
where $\mc{L}_5$ and $\mc{L}_4$ are given in Eqs.~(\ref{L5})
and (\ref{L4}), respectively.
The bulk fields $X_1$, $X_2$, $X_3$ and the brane field $F$ are
auxiliary. Their equations of motion (see Eq.~(\ref{EOM})) are
pure algebraic (contain no derivatives) and can be used to
set these fields to their on-shell values 
(denoted by the ``breve'' accent or the superscript ``on''),
\bea
&& \breve{X}_1 = 0, \qquad 
\breve{X}_2 = 0, \qquad 
\breve{F} = 0 \\
\label{x3on}
&& \breve{X}_3 = -\frac{1}{2}\phi\phi^\ast\da(z).
\eea
We would like to see if the on-shell expressions for 
$\mc{L}_5$ and $\mc{L}_4$ are supersymmetric. Instead of
performing an explicit check, we use a short-cut.

Let us separate out terms containing the auxiliary fields,
\bea
&& \mc{L}_5 = \wh{\mc{L}}_5
+\frac{1}{2}(X_1^2+X_2^2+X_3^2) \\
&& \mc{L}_4 = \wh{\mc{L}}_4
+F F^\ast +\frac{1}{2}\phi\phi^\ast X_3 \; .
\eea
The supersymmetry variation of the hatted quantities (containing
no auxiliary fields) commutes with setting the 
auxiliary fields to their on-shell values. Noting also that 
only $\breve{X}_3$ is non-zero, we obtain,
\bea
\da_\eta(\mc{L}_5^\text{on})-(\da_\eta\mc{L}_5)^\text{on}
&=& \breve{X}_3
\left[\da_\eta\breve{X}_3-(\da_\eta X_3)^\text{on}\right] \\
\da_\eta(\mc{L}_4^\text{on})-(\da_\eta\mc{L}_4)^\text{on}
&=& \frac{1}{2}\phi\phi^\ast
\left[\da_\eta\breve{X}_3-(\da_\eta X_3)^\text{on}\right].
\eea
We know that
\bea
\da_\eta\mc{L}_5=0, \qquad
\da_\eta\mc{L}_4=0 \; ,
\eea
omitting total $\p_m$ and $\p_5$
derivatives, which are both irrelevant in the orbifold picture.
The expression in the square brackets is proportional to
the $\la_1$ equation of motion (see Eq.~(\ref{EOM})),
\be
\da_\eta\breve{X}_3-(\da_\eta X_3)^\text{on}
=\eta_1\left\{
i\p_5\la_2+\si^m\p_m\labar_1-\frac{1}{\st}\phi^\ast\psi\da(z)
\right\}+h.c. 
\ee
Since we are not allowed to use the equations of motion in
checking supersymmetry, we conclude that the on-shell Lagrangians
are not (separately) supersymmetric,
\bea
\da_\eta(\mc{L}_5^\text{on}) \neq 0, \qquad
\da_\eta(\mc{L}_4^\text{on}) \neq 0 \; .
\eea
On the other hand, for the total bulk-plus-brane Lagrangian
we have
\bea
\da_\eta(\mc{L}^\text{on})-(\da_\eta\mc{L})^\text{on}
= \left(\breve{X}_3+\frac{1}{2}\phi\phi^\ast\da(z)\right)
\left[\da_\eta\breve{X}_3-(\da_\eta X_3)^\text{on}\right],
\eea
which does vanish due to Eq.~(\ref{x3on}).
Therefore, the total Lagrangian is
supersymmetric,
\bea
\da_\eta(\mc{L}^\text{on}) =0 \; ,
\eea
and this does not rely on using any boundary conditions.

\subsection{On-shell in the boundary picture}
The total bulk-plus-boundary action is
\bea
S=\int_{\mc{M}}\mc{L}_5{}^\prime
+\frac{1}{2}\int_{\p\mc{M}}\mc{L}_4{}^{\prime} \; .
\eea
Omitting total $\p_m$ derivatives, but keeping total $\p_5$
derivatives (essential in the boundary picture), we have,
\bea
\mc{L}_5{}^\prime = \mc{L}_5
+\p_5\left(-\Phi D-\frac{1}{2}(\la_1\la_2+h.c.)\right), \qquad
\mc{L}_4{}^\prime = \mc{L}_4 \; .
\eea
There is only a slight modification to the expressions obtained
in the previous subsection,
\bea
\da_\eta({\mc{L}_5{}^\prime}^\text{on})
-(\da_\eta\mc{L}_5{}^\prime)^\text{on}
&=& \breve{X}_3
\left[\da_\eta\breve{X}_3-(\da_\eta X_3)^\text{on}\right] 
-\p_5\left(\Phi
\left[\da_\eta\breve{X}_3-(\da_\eta X_3)^\text{on}\right]
\right) 
\\
\da_\eta({\mc{L}_4{}^\prime}^\text{on})
-(\da_\eta\mc{L}_4{}^\prime)^\text{on}
&=& \frac{1}{2}\phi\phi^\ast
\left[\da_\eta\breve{X}_3-(\da_\eta X_3)^\text{on}\right].
\eea
But in the boundary picture, instead of a single auxiliary
field equation (\ref{x3on}), the variational principle for
arbitrary $\da X_3$ in the bulk and on the boundary
produces a bulk equation of motion
\emph{and} a boundary condition,
\bea
\label{auxbc}
\fbox{\;$\dst
\breve{X}_3 =0 \qquad \oplus \qquad
\Phi \beq  -\frac{1}{4}\phi\phi^\ast.
$}
\eea
This is so because the terms involving $X_3$ in the 
bulk-plus-boundary action are
\bea
S=\int_\mc{M}\frac{1}{2}X_3^2
+\int_{\p\mc{M}}(X_3-\p_5\Phi)(\Phi+\frac{1}{4}\phi\phi^\ast)+\dots
\eea

We also know that
\bea
\da_\eta\mc{L}_5{}^\prime=0, \qquad
\da_\eta\mc{L}_4{}^\prime=0
\eea
(now omitting only total $\p_m$ derivatives). Therefore,
we find
\bea
\da_\eta({\mc{L}_5{}^\prime}^\text{on}) &=& \p_5
\left[\Phi (\da_\eta X_3)^\text{on}\right] \\
\da_\eta({\mc{L}_4{}^\prime}^\text{on})
&=& -\frac{1}{2}\phi\phi^\ast(\da_\eta X_3)^\text{on}.
\eea
Combining these expressions, we obtain
\bea
\da_\eta(S^\text{on}) =\int_{\p\mc{M}}
\left\{
-\left(\Phi+\frac{1}{4}\phi\phi^\ast\right)
(\da_\eta X_3)^\text{on}
\right\}.
\eea
Therefore, the total bulk-plus-boundary action, restricted
to on-shell field content, is supersymmetric only if we
use the boundary condition $B_1(\Phi)$, Eq.~(\ref{bc1}),
\bea
\label{usebc}
2\Phi \beq \dst -\frac{1}{2}\phi\phi^\ast 
\qquad \Longrightarrow \qquad
\da_\eta(S^\text{on}) =0 \; .
\eea
But this is exactly the boundary condition which comes 
as a part of the auxiliary field equation for $X_3$, 
Eq.~(\ref{auxbc}). Thus, using this (``auxiliary'') 
boundary condition is just a part of the going on-shell
procedure!

Note that in the boundary picture the boundary condition
$B_1(\Phi)$ also comes as a factor with the general variation 
$\da D$ (and thus could also be called $B_1(D)$), and that on-shell
$D=-\p_5\Phi$. Variations of $\Phi$ and $\p_5\Phi$ on the
boundary are independent. Our Gibbons-Hawking-like $Y$-term makes
only the variation of $\p_5\Phi$ appear on the boundary.

\subsection{On-shell closure of the supersymmetry algebra}
It is well-known that on-shell the supersymmetry algebra closes
only up to equations of motion. This is true in our case as well,
as one can explicitly check. But one should remember that in
the orbifold picture the equations of motion contain $\da(z)$
singularities. And this brings about one important issue.

Among all on-shell fields,
only $\la_1$ has a singular term in its on-shell
supersymmetry transformation. Indeed, off-shell we have
\bea
\da_\eta\la_1=(\si^{mn}F_{mn}-i\p_5\Phi+i X_3)\eta_1 \; ,
\eea
which on-shell (in the orbifold picture) becomes
\bea
\da_\eta\la_1=\si^{mn}\eta_1 F_{mn}
-i\left[\p_5\Phi+\frac{1}{2}\phi\phi^\ast\da(z)\right]\eta_1 \; .
\eea
The commutator of two supersymmetry transformations on $\la_1$ gives
\bea
[\da_\xi, \da_\eta]\la_1 =
-2i U^m_{\xi\eta}\p_m\la_1
+\frac{1}{2}U^m_{\xi\eta}(\si_m\ov{E[\la_1]})
-(\xi\si^{mn}\eta)(\si_{mn}E[\la_1]) \; ,
\eea
where $U^m_{\xi\eta}=\xi\si^m\etabar-\eta\si^m\xibar$ ,
and
\bea
E[\la_1]\equiv
-i\si^m\p_m\labar_1+\p_5\la_2
+\frac{i}{\st}\phi^\ast\psi\da(z)=0
\eea
is the equation of motion which comes with $\da\la_1$, 
see Eq.~(\ref{EOM}). Note that without the singular term in
$\da_\eta\la_1$, the commutator closes up to the non-singular
equation of motion.

We can make the following observations regarding the singular
term
\bea
\da^\prime\la_1 =-\frac{i}{2}\phi\phi^\ast\eta_1\da(z) \; :
\eea
\begin{itemize}
\item[1)]
It comes from eliminating the auxiliary field $X_3$.
\item[2)]
It makes the on-shell supersymmetry algebra close up to the 
full (singular) equations of motion.
\item[3)]
It makes $\da_\eta\la_1$ \emph{non-singular} when the boundary
condition $B_1(\Phi)$ 
(which fixes the jump of $\Phi$ across the brane) 
is taken into account.
\end{itemize}
The last observation provides a ``rule of thumb'' procedure 
for the appropriate modification of the supersymmetry 
transformations in the orbifold picture:
\begin{enumerate}
\item
Identify all terms with $\p_5$ acting on the odd fields.
\item
Find the corresponding natural boundary conditions.
\item
Add singular terms that make the modified supersymmetry
transformation \emph{non-singular} when the boundary conditions 
are taken into account.
\end{enumerate}
This approach was already used in Ref.~\cite{bb1}.

\subsection{The full square structure}
The on-shell Lagrangian in the orbifold picture contains
a $\da(z)^2$ term which comes from eliminating the $X_3$ 
auxiliary field. It turns out that this term combines with
other terms into a full square. (This structure appears all
the time in the orbifold constructions. It was first noticed
by Horava \cite{hor}. We arrive to it also in Ref.~\cite{my2}.) 
Indeed,
\bea
\mc{L} &=& \mc{L}_5+\mc{L}_4\da(z) 
\nn\\
&=& -\frac{1}{2}(\p_5\Phi)^2+\frac{1}{2}X_3^2
+\frac{1}{2}\phi\phi^\ast(X_3-\p_5\Phi)\da(z)+\dots
\nn\\
&=& -\frac{1}{2}\Big[
\p_5\Phi+\frac{1}{2}\phi\phi^\ast\da(z)
\Big]^2+\dots
\eea
The complete on-shell Lagrangian in the orbifold picture
can be written as follows,
\bea
\mc{L} &=& -\frac{1}{4}F_{mn}F^{mn}-\frac{1}{2}\p_m\Phi\p^m\Phi
-\Big[ \frac{i}{2}\la_1\si^m\p_m\labar_1
+\frac{i}{2}\la_2\si^m\p_m\labar_2 +h.c. \Big]
\nn\\
&&-\frac{1}{2}F_{m5}F^{m5}
-\frac{1}{2}\Big[ \p_5\Phi+\frac{1}{2}\phi\phi^\ast\da(z) \Big]^2
+\Big\{ \la_1\big[ 
   \p_5\la_2+\frac{i}{\sqrt2}\phi^\ast\phi\da(z)
             \big]+h.c. \Big\}
\nn\\
&&-\Big[\mc{D}_m\phi\mc{D}^m\phi^\ast
     +i\psibar\sibar^m\mc{D}_m\psi\Big]\da(z) \; .
\eea
We see that the combination 
$\p_5\Phi+\frac{1}{2}\phi\phi^\ast\da(z)$
appears both in the on-shell supersymmetry transformations and in
the on-shell Lagrangian. The special property of this combination
is that it is \emph{non-singular} when the boundary condition
for $\Phi$ is used. 
(Its analog in the five-dimensional supergravity \cite{my2}
is $F_{m5}+2J_m\da(z)$.)

\subsection{On-shell closure of the boundary conditions}
Our natural boundary condition, Eq.~(\ref{bcsf}), is 
\bea
\label{Qbc}
2\p_5{\bf V}-({\bf\Phi_2}+{\bf\Phi_2}^\dagger) \beq
\frac{1}{2}{\bf\Phi}^\dagger e^{\bf V}{\bf\Phi} \; .
\eea
In components, there are six boundary conditions: Eqs.~(\ref{bc1})
and (\ref{bc2}). Off-shell they transform into each other
under the ($\eta=\eta_1$) supersymmetry transformations and
thus form a closed orbit (see also Appendix~\ref{app-bcs}).
Once we go on-shell, however, the
boundary conditions are \emph{no longer closed} under supersymmetry.
Indeed, setting $F=0$ in $B_2(A_m)$ breaks the orbit, since
$\da F=0$ requires using the brane fermionic equation of motion
(see Eqs.~(\ref{strF}) and (\ref{EOM})). 
Therefore, we need to use equations
of motion to close the orbit of boundary conditions 
under supersymmetry on-shell
(same as for the on-shell closure of the supersymmetry algebra!).

\subsection{The role of ``odd $=0$'' boundary conditions}

When the bulk-brane coupling constant $Q$ is written explicitly
(see footnote \ref{footQ}), 
it multiplies the right-hand side
of Eq.~(\ref{Qbc}). Setting it to zero, gives the ``odd $=0$''
boundary conditions,
\bea
2\p_5{\bf V}-({\bf\Phi_2}+{\bf\Phi_2}^\dagger) \beq 0 \; ,
\eea
which in components give
\bea
F_{m5}\beq 0, \quad \Phi\beq 0, \quad \la_2\beq 0 \quad
\p_5\la_1\beq 0, \quad X_{12}\beq 0, \quad \p_5 D \beq 0 \; .
\eea
(It is interesting to note that they 
are closed under supersymmetry both off- and on-shell.)
These boundary conditions were used in the original paper
by Mirabelli and Peskin \cite{mp} as 
\emph{free field boundary conditions}, for example,
to derive the $k^5$ momentum quantization 
(see Eq.~(26) in Ref.~\cite{mp}). The coupling constant $Q$
(or its non-abelian analog $g t^A_{ab}$) is then reintroduced
along the standard logic of the perturbation theory.

If one assumes, instead, that the ``odd $=0$'' boundary conditions 
are the \emph{exact} boundary conditions for the action with $Q\neq 0$,
one would run into an inconsistency.
These boundary conditions do not satisfy Eq.~(\ref{bcsf})
and thus do not lead to the vanishing of the general variation of
the action. It vanishes then only when the variations of the fields
on the boundary are restricted to vanish themselves
($\da{\bf V}\beq 0$, ``keeping $\bf V$ fixed on the boundary''
\`a la Dirichlet boundary condition). But this implies fixing 
$\la_1$ and $\p_5\la_1$ on the boundary at the same time, which gives
an overdetermined boundary value problem!
Also, since Eq.~(\ref{usebc}) does not hold,
the on-shell bulk-plus-boundary action is \emph{not} 
supersymmetric with these boundary conditions.

Therefore, the ``odd $=0$'' boundary conditions can only be
treated as the free field boundary conditions, that is as
a starting point in the perturbative calculations.


\section{From boundary to orbifold picture}
\label{sec-mp6}
In order to better understand the role of the $Y$-term, 
Eq.~(\ref{Yterm}), let us
discuss the transition from the boundary to the orbifold 
picture.\footnote{For an earlier discussion of the relationship
between the orbifold and boundary pictures see, e.g., 
Ref.~\cite{lama1}.}

In the boundary picture we have the following total action,
\bea
S^{(+)}=\int_{\mc{M}_+}\mc{L}_5+\int_{\p\mc{M}_+}Y^{(+)}
+\frac{1}{2}\int_{\p\mc{M}_+}\mc{L}_4 \; ,
\eea
where $\mc{M}_+=\mathbb{R}^{1,3}\times[0, +\infty)$.
The orbifold is, essentially, a union of two domains with
boundary. The reflection of $\mc{M}_+$ is 
$\mc{M}_-=\mathbb{R}^{1,3}\times(-\infty, 0]$. Its total
bulk-plus-boundary action is
\bea
S^{(-)}=\int_{\mc{M}_-}\mc{L}_5-\int_{\p\mc{M}_-}Y^{(-)}
+\frac{1}{2}\int_{\p\mc{M}_-}\mc{L}_4 \; ,
\eea
The choice of signs is easy to understand. First, we have
\bea
\int_{\mc{M}_+}\p_5 K^5 =
\int_{\p\mc{M}_+}(-K^5), \qquad
\int_{\mc{M}_-}\p_5 K^5 =
\int_{\p\mc{M}_-}(+K^5) \; ,
\eea
which says that the signs of $Y^{(-)}$ and $Y^{(+)}$ 
relative to $\mc{L}_5$ should be opposite.
Second, in the orbifold picture the $Y$-term is odd
(since $\la_2$ and $\Phi$ are odd). Therefore, to get the
correct boundary conditions, the signs of $Y^{(-)}$ and $Y^{(+)}$
relative to $\mc{L}_4$ should be opposite.

The boundaries of $\mc{M}_+$ and $\mc{M}_-$ coincide, so
we denote 
\bea
\Si=\p\mc{M}_+=\p\mc{M}_- \; .
\eea
Since the $Y$-term is odd,
$Y^{(-)}=-Y^{(+)}$, the sum of the two bulk-plus-boundary
actions is
\bea
S=\int_{\mc{M}_+ \cup \mc{M}_-}\mc{L}_5
+\int_\Si 2Y^{(+)}
+\int_\Si \mc{L}_4 \; .
\eea
We now want to show that this equals to our bulk-plus-brane
action,
\bea
S =
\int_{\mc{M}_5} \mc{L}_5 +
\int_\Si \mc{L}_4 =
\int_{\mc{M}_5} \Big\{ \mc{L}_5 + \mc{L}_4\da(z) \Big\} \; ,
\eea
where $\mc{M}_5=\mathbb{R}^{1,4}=
\mathbb{R}^{1,3}\times(-\infty, +\infty)$.
This is equivalent to showing that the $Y$-term matches
onto brane-localized terms produced by the bulk Lagrangian
$\mc{L}_5$.

To do this, it helps to represent odd fields (in our case,
$\Phi$ and $\la_2$) as follows,
\bea
\Phi(x,z)=\ep(z)\Phi^{(+)}(x,|z|) \; ,
\eea
where $\ep(z)={\rm{sgn}}(z)=\pm 1$ on $\mc{M}_\pm$.
For the $\p_5$ derivative of an odd field, we then have 
\bea
\label{p5ep}
\p_5\Phi=\ep\p_5\left[ \Phi^{(+)} \right] 
+\Phi^{(+)}\ep^\prime(z) =
(\p_5\Phi)^{(+)} +2\Phi^{(+)}\da(z) \; .
\eea
(The superscript $(+)$ means ``evaluated on the 
$\mc{M}_+$ side''.) This allows us to separate out 
the $\Si$-localized terms in $\mc{L}_5$. 

The relevant terms in the bulk Lagrangian are
\bea
\mc{L}_5=\frac{1}{2}X_3^2-\frac{1}{2}(\p_5\Phi)^2
+\frac{1}{2}(\la_1\p_5\la_2+h.c.)+\dots
\eea
Using Eq.~(\ref{p5ep}), we find
\bea
\mc{L}_5=\frac{1}{2}X_3^2-2[\Phi^{(+)}\da(z)]^2
-2(\Phi\p_5\Phi)^{(+)}\da(z)
+(\la_1\la_2^{(+)}+h.c.)\da(z)+\dots
\eea
This is to be compared with
\bea
2Y^{(+)}\da(z)=2[\Phi X_3]^{(+)}\da(z)
-2(\Phi\p_5\Phi)^{(+)}\da(z)
+(\la_1\la_2^{(+)}+h.c.)\da(z) \; .
\eea
We see that terms \emph{without} $X_3$ and $\da(z)^2$ do match!
The remaining terms appear to match only \emph{on-shell}. Indeed,
on-shell we have
\bea
X_3 =-\frac{1}{2}\phi\phi^\ast\da(z), \qquad
X_3^{(+)}=0, \qquad 2\Phi^{(+)}=-\frac{1}{2}\phi\phi^\ast
\eea
(including the ``auxiliary boundary condition,'' 
Eq.~(\ref{auxbc})), which implies
\bea
\frac{1}{2}X_3^2-2[\Phi^{(+)}\da(z)]^2 =0, \qquad
[\Phi X_3]^{(+)}=0 \; .
\eea
We conclude, therefore, that the $Y$-term matches onto 
\emph{regular singularities} (just $\da(z)$)
of the bulk Lagrangian, whereas
higher order singularities (in our case $\da(z)^2$) are
taken care of by the auxiliary fields after going on-shell.
In other words, only the on-shell part of the $Y$-term
can be derived from the comparison with the brane-localized
terms produced by the bulk Lagrangian.

\section{Summary and Conclusions}
In this paper we discussed the Mirabelli and Peskin model \cite{mp}
in various settings: in the orbifold and boundary pictures,\footnote{
The physics described by both pictures is guaranteed to be the same
since the boundary conditions on the fundamental domain are identical.
One advantage of the boundary picture is a complete removal of all 
ambiguities related to products of distributions.}
in components and in the superfield formulation, off-shell
and on-shell. 

We showed that the boundary picture requires 
introduction of the $Y$-term (Gibbons-Hawking-like term),
which is necessary for supersymmetry and allows us to derive
natural (Neumann-like) boundary conditions via the standard
application of the variational principle. We found that the
$Y$-term arises naturally in the ($D=4$, $N=1$) superfield 
formulation of the ($D=5$, $N=2$) model. 

We demonstrated that, in the orbifold picture, ($N=1$) supersymmetry 
does not require the use of any boundary conditions both
off- and on-shell. In the boundary picture, however, supersymmetry
of the total action requires the use of one boundary condition:
the ``auxiliary boundary condition,'' Eq.~(\ref{auxbc}) 
(the one which comes as a part of the auxiliary equations of motion).
This boundary condition is also one of the natural boundary
conditions. 

We showed that
the natural boundary conditions form a closed orbit under ($N=1$)
supersymmetry and can be put in a superfield form.
We can identify the ``boundary condition superfield'' 
(see Appendix \ref{app-bcs}).
The natural boundary conditions reduce to 
the ``odd=0'' boundary conditions only in the
absence of coupling to the brane-localized matter.

We also saw what modifications to the supersymmetry transformations
are necessary in the orbifold picture, and confirmed that the $\da(z)^2$
terms fit into the full square structure in the Lagrangian.
We found that the $Y$-term matches onto regular singularities
of the bulk Lagrangian, but that higher order singularities 
(like $\da(z)^2$) are taken care of only on-shell.

The detailed analysis of this simple model 
serves as a basis for the analysis in Ref.~\cite{my2},
where we discuss five-dimensional (on-shell) supergravity 
on a manifold with boundary. 

\acknowledgments
I would like to thank
Jonathan Bagger for helpful discussions and critical reading of
this manuscript. I would also like to thank Fabio Zwirner for
his questions on the closure of boundary conditions in 
Ref.~\cite{bb1} under supersymmetry, which were a part of the
motivation for this work.
This work was supported in part by the National Science Foundation, 
grant NSF-PHY-0401513.


\appendix
\section{Details of the component formulation}
Here we collect various technical details, which were only
briefly mentioned in the body of the paper.
The basic conventions are as in Ref.~\cite{bb1}.

\subsection{Variation of the bulk Lagrangian}
\label{app-var}

Under a general variation of the fields, the bulk Lagrangian
varies as follows,
\bea
\label{gvarL5}
\da\mc{L}_5 &=& 
\da A_N(\p_M F^{MN})+\da\Phi(\p_M\p^M\Phi)
+X_a\da X_a 
\nn\\[5pt]
&&-i\da\wt\La^i\Ga^M\p_M\La_i
+\p_M K^M,
\eea
where
\bea
\label{KM}
K^M=-F^{MN}\da A_N-\da\Phi\p^M\Phi
-\frac{i}{2}\wt\La^i\Ga^M\da\La^i \; .
\eea
The variation of the Lagrangian under the supersymmetry 
transformations, Eq.~(\ref{MPsusytr}), 
prefers a different separation of the total
derivative term,
\bea
\label{svarL5}
\da_\Eta\mc{L}_5 &=& 
\da_\Eta A_N(\p_M F^{MN})+\da_\Eta\Phi(\p_M\p^M \Phi)
+X_a\da_\Eta X_a
\nn\\[5pt]
&&-i\wt\La^i\Ga^M\p_M\da^\prime_\Eta\La_i
-i\da^{\prime\prime}_\Eta\wt\La^i\Ga^M\p_M\La_i
+\p_M \wt K^M ,
\eea
where
\bea
\label{wtKM}
\wt K^M=
-F^{MN}\da_\Eta A_N-\da_\Eta\Phi\p^M\Phi
-\frac{i}{2}\da^\prime_\Eta\wt\La^i\Ga^M\La_i
-\frac{i}{2}\wt\La^i\Ga^M\da^{\prime\prime}_\Eta\La_i
\eea
and we used the following split in the supersymmetry
transformation of the gaugino,
\bea
\da_\Eta\La_i &=& \da^\prime_\Eta\La_i+\da^{\prime\prime}_\Eta\La_i \\[5pt]
&&\da^\prime_\Eta\La_i=(\Si^{MN}F_{MN}+\Ga^M\p_M\Phi)\Eta_i, \quad
\da^{\prime\prime}_\Eta\la_i=X_a(\si_a)_i{}^j\Eta_j \; .
\eea
It is easy to check that terms outside the total derivative
cancel, thus giving
\bea
\da_\Eta\mc{L}_5 &=& \p_M \wt K^M .
\eea

\subsection{Two-component spinor notation}
\label{app-comp}
Making the $M=\{m,5\}$ split in the supersymmetry transformations
and rewriting them in terms of the two-component Weyl spinors, 
we find
\bea
\ba{lcl}
\da_\Eta A^m &=& i(\eta_1\si^m\labar_1+\eta_2\si^m\labar_2)+h.c.\\[5pt]
\da_\Eta A^5 &=& -\eta_1\la_2+\eta_2\la_1+h.c.\\[5pt]
\da_\Eta\Phi &=& i(-\eta_1\la_2+\eta_2\la_1)+h.c.\\[5pt]
\da_\Eta\la_1 &=& (\si^{mn}F_{mn}-i\p_5\Phi+i X_3)\eta_1
+(i F_{m5}+\p_m\Phi)\si^m\etabar_2-i X_{12}^\ast\eta_2\\[5pt]
\da_\Eta\la_2 &=& -(i F_{m5}+\p_m\Phi)\si^m\etabar_1-i X_{12}\eta_1
+(\si^{mn}F_{mn}-i\p_5\Phi-i X_3)\eta_2\\[5pt]
\da_\Eta X_{12} &=& 
2\etabar_1(i\p_5\labar_1-\sibar^m\p_m\la_2)
+2\eta_2(i\p_5\la_2+\si^m\p_m\labar_1) \\[5pt]
\da_\Eta X_3 &=& 
-\eta_1(i\p_5\la_2+\si^m\p_m\labar_1)
-\eta_2(i\p_5\la_1-\si^m\p_m\labar_2) +h.c. \; ,
\ea
\nn\\
\eea
where we defined $X_{12}=X_1+i X_2$. We used the following
relation between the symplectic-Majorana spinor $\La_i$ and the
pair $(\la_1, \la_2)$,
\bea
\La_1 = -\La^2 = \binom{\la_1}{\labar_2}, \quad
\La_2 =  \La^1 = \binom{-\la_2}{\labar_1} .
\eea
The same relation holds between $\Eta_i$ and $(\eta_1, \eta_2)$.
The Majorana conjugation, denoted by a tilde, gives
\be
\wt\La_1=(-\la_1, \labar_2) \; .
\ee

\subsection{$SU(2)$ rotation}
\label{app-su2}
The bulk Lagrangian is invariant
(and the bulk supersymmetry transformations are covariant)
under a global $SU(2)$ rotation,
\bea
\La_i^\prime = \Uij\La_j \; , \qquad
X_a{}^\prime \si_a=U(X_a \si_a)U^\dagger \; ,
\eea
where $U\in SU(2)$ is a constant matrix, and $\si_a$ are
the Pauli matrices such that
\bea
X_a\si_a=\left(\bma X_3 & X_{12}^\ast \\[3pt]
                   X_{12} & -X_3 \ema\right).
\eea
A particularly useful $SU(2)$ rotation is 
\bea
\la_1^\prime=\frac{\la_1-\al^\ast\la_2}{\sqrt{1+\al\al^\ast}}, \qquad
\la_2^\prime=\frac{\al\la_1+\la_2}{\sqrt{1+\al\al^\ast}} 
\eea
(same for $(\eta_1, \eta_2)$),
accompanied by
\bea
X_{12}{}^\prime &=& \frac{X_{12}-\al^2 X_{12}^\ast -2\al X_3}
{1+\al\al^\ast} \\[5pt]
X_3{}^\prime &=& \frac{\al X_{12}^\ast+\al^\ast X_{12}+(1-\al\al^\ast)X_3}
{1+\al\al^\ast} \; .
\eea
The inverse transformation is obtained by changing the sign of $\al$.
In particular,
\bea
\eta_2=\frac{-\al\eta_1^\prime+\eta_2^\prime}{\sqrt{1+\al\al^\ast}} \; ,
\eea
so that 
\bea
\eta_2=0 \qquad \Rightarrow \qquad
\eta_2^\prime =\al\eta_1^\prime \; .
\eea

\subsection{$N=1$ supersymmetry and the gauge transformation}
\label{app-n1}

Under $\eta=\eta_1$ supersymmetry, the five-dimensional gauge
supermultiplet splits into two four-dimensional 
supermultiplets.
These are a gauge multiplet (in the WZ gauge)
\bea
\label{def1}
v_m=A_m, \quad \la=\la_1, \quad D=X_3-\p_5\Phi \; ,
\eea
and a chiral multiplet
\bea
\label{def2}
\phi_2=\Phi+i A_5, \quad \psi_2=-i\st\la_2, \quad F_2=-X_{12} \; .
\eea
The definitions lead to the standard transformation laws for
the gauge multiplet,
\bea
\label{strV}
\ba{lcl}
\da_\eta v_m &=& i\eta\si_m\labar+h.c.\\[5pt]
\da_\eta\la &=& \si^{mn}\eta v_{mn}+i\eta D\\[5pt]
\da_\eta D &=& -\eta\si^m\p_m\labar+h.c. \; ,
\ea
\eea
as well as to the following supersymmetry transformations 
for the chiral multiplet,
\bea
\label{strF2}
\ba{lcl}
\da_\eta\phi_2 &=& \st\eta\psi_2\\[5pt]
\da_\eta\psi_2 &=& i\st\si^m\etabar\p_m\phi_2+\st\eta F_2
+\st\si^m\etabar\p_5 v_m\\[5pt]
\da_\eta F_2 &=& i\st\etabar\sibar^m\p_m\psi_2-2i\etabar\p_5\labar \; .
\ea
\eea
The latter differ from the standard transformations by extra
terms involving $\p_5$. One can check that the supersymmetry
algebra closes up to a gauge transformation,
\bea
\label{algebra}
[\da_\xi,\da_\eta]=-2i U^m_{\xi\eta}\p_m+\da_u \; ,
\eea
where $U^m_{\xi\eta}=\xi\si^m\etabar-\eta\si^m\xibar$ and
the gauge transformation is non-zero only on $v_m$ and $\phi_2$,
\bea
\label{gtr1}
\da_u v_m=\p_m u, \quad 
\da_u\phi_2=i\p_5 u \; ,
\eea
with $ u=2i U^m_{\xi\eta}v_m$. This is just
the $U(1)$ gauge transformation $\da A_M=\p_M u$.

The chiral multiplet $(\phi, \psi, F)$, living on the brane,
has the standard supersymmetry transformations
(note the dependence on $v_m$ and $\la$ from the vector multiplet),
\bea
\label{strF}
\ba{lcl}
\da_\eta\phi &=& \st\eta\psi\\[5pt]
\da_\eta\psi &=& i\st\si^m\etabar\cd_m\phi+\st\eta F\\[5pt]
\da_\eta F &=& i\st\etabar\sibar^m\cd_m\psi 
+i\phi\etabar\labar \; ,
\ea
\eea
where
\bea
\cd_m=\p_m+\frac{i}{2}v_m
\eea
when acting on $(\phi, \psi, F)$. This is the gauge covariant
derivative, corresponding to the following $U(1)$ transformation
of the chiral multiplet,
\bea
\label{gtr2}
\da_u(\phi, \psi, F)=-\frac{i}{2} u (\phi, \psi, F) \; .
\eea
The algebra (\ref{algebra}) holds for this multiplet
as well.

\subsection{Orbifold equations of motion}
\label{app-orb}
In the orbifold picture the total Lagrangian is
$\mc{L}=\mc{L}_5+\da(z)\mc{L}_4$.
For a general variation of the fields, we find
\bea
\da\mc{L}_5 &=&
\da A_n(\p_m F^{mn}+\p_5 F^{5n})+\da A_5(\p_m F^{m5})
+\da\Phi(\p_m\p^m\Phi+\p_5\p_5\Phi)+X_a\da X_a
\nn\\[5pt]
&&-\big[\da\la_1(i\si^m\p_m\labar_1-\p_5\la_2)
+\da\la_2(i\si^m\p_m\labar_2+\p_5\la_1)+h.c.\big]\\
\da\mc{L}_4 &=&
\da\phi\big(
\cd_m\cd^m\phi^\ast+\frac{1}{2}\phi^\ast D
-\frac{i}{\st}\labar\psibar
\big)
+\da\psi\big(
-i\si^m\cd_m\psibar+\frac{i}{\st}\phi^\ast\la
\big)
+F^\ast\da F \nn\\
&&+\da v_m\big(
-\frac{i}{2}\phi\cd^m\phi^\ast
-\frac{1}{4}\psi\si^m\psibar
\big)
+\da\la\big( \frac{i}{\st}\phi^\ast\psi \big)
+\da D \big( \frac{1}{4}\phi\phi^\ast \big)+h.c.
\eea
(Total $\p_m$ and $\p_5$ derivatives have been omitted here.)
Using the definitions in Eq.~(\ref{def1}) and rewriting
\bea
\da D\phi\phi^\ast\da(z)=
\da X_3\phi\phi^\ast\da(z)+\da\Phi\p_5(\phi\phi^\ast\da(z)) \; ,
\eea
we find the following equations of motion,
\bea
\label{EOM}
\ba{lcrcl}
E(A_m) &:& \dst
\p_n F^{nm}-\p_5 F^{m5}-\left(
\frac{i}{2}(\phi\cd^m\phi^\ast-\phi^\ast\cd^m\phi)
+\frac{1}{2}\psi\si^m\psibar\right)\da(z) &=& 0\\[8pt]
E(A_5) &:& \dst \p_m F^{m5} &=& 0\\[8pt]
E(\Phi) &:& \dst \p_m\p^m\Phi
+\p_5(\p_5\Phi+\frac{1}{2}\phi\phi^\ast\da(z)) &=& 0\\[8pt]
E(\la_1) &:& \dst -i\si^m\p_m\labar+\p_5\la_2
+\frac{i}{\st}\phi^\ast\psi\da(z) &=& 0\\[10pt]
E(\la_2) &:& \dst -i\si^m\p_m\labar_2-\p_5\la &=& 0\\[8pt]
E(X_{12}) &:& \dst X_{12}^\ast &=& 0\\[5pt]
E(X_3) &:& \dst X_3+\frac{1}{2}\phi\phi^\ast\da(z) &=& 0\\[8pt]
E(\phi) &:& \dst \cd_m\cd^m\phi^\ast+\frac{1}{2}\phi^\ast D
-\frac{i}{\st}\labar\psibar &=& 0\\[10pt]
E(\psi) &:& \dst -i\si^m\cd_m\psibar
+\frac{i}{\st}\phi^\ast\la &=& 0\\[8pt]
E(F) &:& \dst F^\ast &=& 0 \; .
\ea
\nn\\
\eea

\section{Superfields}
We follow the conventions of Ref.~\cite{wb}. 
See also Ref.~\cite{kuz}.

\subsection{Supersymmetry and gauge transformations}
\label{app-sf}

The vector and chiral superfields have the following component
expansions,
\bea
&&{\bf V}= -i\ta\si^m\tabar v_m +i\ta^2\tabar\labar-i\tabar{}^2\ta\la
+\frac{1}{2}\ta^2\tabar{}^2 D \\
&&{\bf\Phi}=\phi+i\ta\si^m\tabar\p_m\phi
+\frac{1}{4}\ta^2\tabar{}^2\p_m\p^m\phi
+\st\ta\psi+\frac{i}{\st}\ta^2\tabar\sibar^m\p_m\psi
+\ta^2 F \; .
\eea
The chiral superfields can be more conveniently
written in terms of the ``$y$ coordinates'' ($y^m=x^m+\ta\si^m\tabar$),
\bea
{\bf\Phi}(y)=\phi(y)+\st\ta\psi(y)+\ta^2 F(y) \; .
\eea
The field strength $\bf W$ of the vector superfield $\bf V$ is
a chiral spinor superfield (its lowest component is a spinor),
which in the $y$-coordinates has the following form,
\bea
{\bf W}(y)=-i\la+\ta D-i\si^{mn}\ta v_{mn}+\ta^2(\si^m\p_m\labar) \; ,
\eea
where $v_{mn}=\p_m v_n-\p_n v_m$.

The supersymmetry transformations in Eqs.~(\ref{strV}), (\ref{strF2})
and (\ref{strF}) can be written in the following superfield form,
\bea
\label{supertr}
\ba{lcl}
\da_\eta{\bf V} &=& (\eta Q+\etabar\ov Q){\bf V}
+{\bf\Phi_\eta}+{\bf\Phi_\eta}^\dagger \\[3pt]
\da_\eta{\bf\Phi} &=& (\eta Q+\etabar\ov Q){\bf\Phi}
-{\bf\Phi_\eta}{\bf\Phi}\\[3pt]
\da_\eta{\bf\Phi_2} &=& (\eta Q+\etabar\ov Q){\bf\Phi_2}
+2\p_5{\bf\Phi_\eta} \; ,
\ea
\eea
where $\bf\Phi_\eta$ is a chiral superfield given by
\bea
{\bf\Phi_\eta}(y)=\st\ta(\frac{1}{\st}\si^m\etabar v_m)
+\ta^2(-i\etabar\labar) \; .
\eea
It describes a compensating supergauge transformation necessary
to keep $\bf V$ in the WZ gauge. From this we deduce that the
supergauge transformation for all the superfields is given by
\bea
\bf
\da V=\La+\La^\dagger, \quad
\da\Phi=-\La\Phi, \quad
\da\Phi_2=2\p_5\La \; .
\eea
The residual gauge transformation, preserving the WZ gauge, 
corresponds to
\bea
-2i{\bf\La}= u(y)= u+i\ta\si^m\tabar\p_m u
+\frac{1}{2}\ta^2\tabar{}^2\p_m\p^m u \; .
\eea
In components this gauge transformation reproduces
Eqs.~(\ref{gtr1}) and (\ref{gtr2}).

\subsection{Component expansions}
\label{app-compexp}
The component expansions for the bulk action are:
\bea
\int d^2\ta {\bf W} {\bf W} +h.c. =
-v_{mn}v^{mn}-2i(\la\si^m\p_m\labar+\labar\sibar^m\p_m\la)
+2D^2 \\
\int d^2\ta d^2\tabar \; {\bf Z}^2 =
-\frac{1}{2}\p_m\phi_2\p^m\phi_2^\ast
+\frac{1}{2}F_2 F_2^\ast
-\frac{i}{4}(\psibar_2\sibar^m\p_m\psi_2+\psi_2\si^m\p_m\psibar_2)
\nn\\
-\frac{1}{2}(\p_5 v_m)(\p_5 v^m)
-\frac{i}{2}(\p_m\phi_2-\p_m\phi_2^\ast)\p_5 v^m
-\frac{1}{2}(\phi_2+\phi_2^\ast)\p_5 D
\nn\\
+\frac{i}{\st}(\psibar_2\p_5\labar-\psi_2\p_5\la)
+\frac{1}{16}\p^m\p_m(\phi_2+\phi_2^\ast)^2 \; .
\eea
The component expansions for the brane action 
and the boundary conditions are:
\bea
\p_5{\bf V} &=& 
-i\tabar{}^2\ta\p_5\la+h.c.
-i\ta\si^m\tabar \p_5 v_m 
+\frac{1}{2}\ta^2\tabar{}^2 \p_5 D 
\\
{\bf\Phi_2}+{\bf\Phi_2}^\dagger &=&
\st\ta\psi_2+\ta^2 F_2 
+\frac{i}{\st}\tabar{}^2\ta\si^m\p_m\psibar_2 +h.c.
\nn\\
&&+(\phi_2+\phi_2^\ast)
+i\ta\si^m\tabar(\p_m\phi_2-\p_m\phi_2^\ast)
+\frac{1}{4}\ta^2\tabar{}^2\p_m\p^m(\phi_2+\phi_2^\ast)
\\
{\bf\Phi}^\dagger e^{\bf V}{\bf\Phi} &=&
\st\ta(\phi^\ast\psi)+\ta^2(\phi^\ast F)\nn\\
&&+\tabar{}^2\ta\left(
-i\la\phi\phi^\ast
+\frac{i}{\st}\si^m(\cd_m\psibar\phi-\psibar\cd_m\phi)
+\st\psi F^\ast\right) +h.c.\nn\\
&&+(\phi\phi^\ast)
+\ta\si^m\tabar\left(
-i(\phi\cd_m\phi^\ast-\phi^\ast\cd_m\phi)
-\psi\si_m\psibar\right)\nn\\
&&+\ta^2\tabar{}^2\left(
\mc{L}_4^r+\frac{1}{4}\p_m\p^m(\phi\phi^\ast)\right) ,
\eea
where $\mc{L}_4^r$ is given in Eq.~(\ref{L4r}).

\subsection{Boundary condition superfield}
\label{app-bcs}

The boundary condition superfield is
\bea
{\bf B} =
2\,\p_5{\bf V}-({\bf\Phi_2}+{\bf\Phi_2}^\dagger)
-\frac{1}{2}{\bf\Phi}^\dagger e^{\bf V}{\bf\Phi} \; .
\eea
Since $\bf B$ is a vector superfield, we can write \cite{wb}
\bea
{\bf B} &=&
i\ta\chi^B
+\frac{i}{2}\ta^2 M^B
-i\ta^2\tabar\la^B
+h.c. \nn\\
&&+\,C^B
-i\ta\si^m\tabar v^B_m
+\frac{1}{2}\ta^2\tabar^2 D^B \; ,
\eea
and identify each component as
\bea
C^B &=& -\frac{1}{2}\phi\phi^\ast -2\Phi \nn\\
\chi^B &=& \frac{i}{\st}\phi^\ast\phi+2\la_2 \nn\\
M^B &=& i(F\phi^\ast-2 X_{12}) \nn\\
v^B_m &=& -\frac{i}{2}(\phi D_m\phi^\ast-\phi^\ast D_m\phi)
-\frac{1}{2}\phi\si_m\psibar-2F_{m5} \nn\\
\la^B &=& -\frac{i}{\st}\si^m\psibar D_m\phi
+\frac{1}{\st}\psi F^\ast-\frac{i}{2}\phi\phi^\ast\la_1
+2i\p_5\la_1 \nn\\
D^B &=& -\mc{L}_4^r+2\p_5 D \; .
\eea
The relation to the boundary conditions, Eqs.~(\ref{bc1}) and (\ref{bc2}),
is as follows,
\bea
\ba{l@{\hspace{40pt}}l@{\hspace{40pt}}l}
C^B: B_1(\Phi) & \chi^B: B_1(\la_1) & M^B: B_2(\la_1) \\[5pt]
v^B_m: B_1(A_m) & \la^B: B_2(A_m) & D^B: B_3(A_m) \; .
\ea
\eea
Using Eq.~(\ref{supertr}), we can easily show that
\bea
\da_\eta{\bf B} &=& (\eta Q+\etabar\ov Q){\bf B} \; ,
\eea
which in components gives (the superscript $B$ is omitted for clarity)
\bea
\da_\eta C &=& i\eta\chi+h.c. \nn\\
\da_\eta \chi &=& \si^m\eta(\p_m C+i v_m)+\eta M \nn\\
\da_\eta M &=& 2\etabar(\labar+\sibar^m\p_m\chi) \nn\\
\da_\eta D &=& -\eta\si^m\p_m\labar+h.c. \nn\\
\da_\eta \la &=& 2\si^{mn}\eta\p_m v_n+i\eta D \nn\\
\da_\eta v_m &=& i\eta\si_m\labar+\eta\p_m\chi +h.c.
\eea
This shows explicitly the structure of supersymmetry variations
in the orbit of boundary conditions (cf. Eq.~(\ref{structure})).



\begin{thebibliography}{99}

\bibitem{hw}
  P.~Horava and E.~Witten,
  ``Eleven-Dimensional Supergravity on a Manifold with Boundary,''
  Nucl.\ Phys.\ B {\bf 475}, 94 (1996)
  [arXiv:hep-th/9603142].

\bibitem{mp}
  E.~A.~Mirabelli and M.~E.~Peskin,
  ``Transmission of supersymmetry breaking from a 4-dimensional boundary,''
  Phys.\ Rev.\ D {\bf 58}, 065002 (1998)
  [arXiv:hep-th/9712214].

\bibitem{rs1}
  L.~Randall and R.~Sundrum,
  ``A large mass hierarchy from a small extra dimension,''
  Phys.\ Rev.\ Lett.\  {\bf 83}, 3370 (1999)
  [arXiv:hep-ph/9905221].

\bibitem{rs2}
  L.~Randall and R.~Sundrum,
  ``An alternative to compactification,''
  Phys.\ Rev.\ Lett.\  {\bf 83}, 4690 (1999)
  [arXiv:hep-th/9906064].

\bibitem{abn}
R.~Altendorfer, J.~Bagger and D.~Nemeschansky,
``Supersymmetric Randall-Sundrum scenario,''
Phys.\ Rev.\ D {\bf 63}, 125025 (2001)
[hep-th/0003117].

\bibitem{gp1}
T.~Gherghetta and A.~Pomarol,
``Bulk fields and supersymmetry in a slice of AdS,''
Nucl.\ Phys.\ B {\bf 586}, 141 (2000)
[hep-ph/0003129].

\bibitem{flp1}
A.~Falkowski, Z.~Lalak and S.~Pokorski,
``Supersymmetrizing branes with bulk in five-dimensional supergravity,''
Phys.\ Lett.\ B {\bf 491}, 172 (2000)
[hep-th/0004093].

\bibitem{bb1}
  J.~Bagger and D.~V.~Belyaev,
  ``Supersymmetric branes with (almost) arbitrary tensions,''
  Phys.\ Rev.\ D {\bf 67}, 025004 (2003)
  [arXiv:hep-th/0206024].

\bibitem{my2}
D.~V.~Belyaev,
``Boundary conditions in supergravity on a manifold with boundary,''
arXiv:hep-th/0509172.

\bibitem{barth}
  N.~H.~Barth,
  ``The Fourth Order Gravitational Action For Manifolds With Boundaries,''
  Class.\ Quant.\ Grav.\  {\bf 2}, 497 (1985).

\bibitem{gibh}
  G.~W.~Gibbons and S.~W.~Hawking,
  ``Action Integrals And Partition Functions In Quantum Gravity,''
  Phys.\ Rev.\ D {\bf 15}, 2752 (1977).

\bibitem{york1}
J.~W.~York, Jr.; 
``Role of conformal three-geometry in the dynamics of gravitation,''
Phys.\ Rev.\ Lett.\  {\bf 28}, 1082 (1972).

\bibitem{york2}
J.~W.~York, Jr.; 
``Boundary terms in the action principles of general relativity,''
Foundations of Physics, {\bf 16}, 249 (1986). 

\bibitem{dive1}
  P.~Di Vecchia, B.~Durhuus, P.~Olesen and J.~L.~Petersen,
  ``Fermionic Strings With Boundary Terms,''
  Nucl.\ Phys.\ B {\bf 207}, 77 (1982).

\bibitem{igarashi}
  Y.~Igarashi,
  ``Supersymmetry And The Casimir Effect Between Plates,''
  Phys.\ Rev.\ D {\bf 30}, 1812 (1984);
  Y.~Igarashi and T.~Nonoyama,
  ``Supergravity And Casimir Energy In A Plane Geometry,''
  Phys.\ Lett.\ B {\bf 161}, 103 (1985);
  Y.~Igarashi and T.~Nonoyama,
  ``Supersymmetry And Reflective Boundary Conditions In Anti-De Sitter
  Spaces,''
  Phys.\ Rev.\ D {\bf 34}, 1928 (1986).


\bibitem{global}
  C.~Albertsson, U.~Lindstrom and M.~Zabzine,
  ``N = 1 supersymmetric sigma model with boundaries. I,''
  Commun.\ Math.\ Phys.\  {\bf 233}, 403 (2003)
  [arXiv:hep-th/0111161];
\newline
  C.~Albertsson, U.~Lindstrom and M.~Zabzine,
  ``N = 1 supersymmetric sigma model with boundaries. II,''
  Nucl.\ Phys.\ B {\bf 678}, 295 (2004)
  [arXiv:hep-th/0202069];
\newline
  I.~V.~Melnikov, M.~R.~Plesser and S.~Rinke,
  ``Supersymmetric boundary conditions for the N = 2 sigma model,''
  arXiv:hep-th/0309223;
\newline
  P.~Koerber, S.~Nevens and A.~Sevrin,
  ``Supersymmetric non-linear sigma-models with boundaries revisited,''
  JHEP {\bf 0311}, 066 (2003)
  [arXiv:hep-th/0309229].

\bibitem{nieu}
  U.~Lindstrom, M.~Rocek and P.~van Nieuwenhuizen,
  ``Consistent boundary conditions for open strings,''
  Nucl.\ Phys.\ B {\bf 662}, 147 (2003)
  [arXiv:hep-th/0211266];
\newline
  P.~van Nieuwenhuizen and D.~V.~Vassilevich,
  ``Consistent boundary conditions for supergravity,''
  arXiv:hep-th/0507172.

\bibitem{agw}
N.~Arkani-Hamed, T.~Gregoire and J.~Wacker,
``Higher dimensional supersymmetry in 4D superspace,''
JHEP {\bf 0203}, 055 (2002)
[arXiv:hep-th/0101233].

\bibitem{heb}
A.~Hebecker,
``5D super Yang-Mills theory in 4-D superspace, superfield brane  operators,
and applications to orbifold GUTs,''
Nucl.\ Phys.\ B {\bf 632}, 101 (2002)
[arXiv:hep-ph/0112230].

\bibitem{csaki1}
  C.~Csaki, C.~Grojean, H.~Murayama, L.~Pilo and J.~Terning,
  ``Gauge theories on an interval: Unitarity without a Higgs,''
  Phys.\ Rev.\ D {\bf 69}, 055006 (2004)
  [arXiv:hep-ph/0305237].

\bibitem{csaki2}
  C.~Csaki, C.~Grojean, J.~Hubisz, Y.~Shirman and J.~Terning,
  ``Fermions on an interval: Quark and lepton masses without a Higgs,''
  Phys.\ Rev.\ D {\bf 70}, 015012 (2004)
  [arXiv:hep-ph/0310355].


\bibitem{hor}
  P.~Horava,
  ``Gluino condensation in strongly coupled heterotic string theory,''
  Phys.\ Rev.\ D {\bf 54}, 7561 (1996)
  [arXiv:hep-th/9608019].


\bibitem{lama1}
  Z.~Lalak and R.~Matyszkiewicz,
  ``Boundary terms in brane worlds,''
  JHEP {\bf 0111}, 027 (2001)
  [arXiv:hep-th/0110141].


\bibitem{wb}
J.~Wess and J.~Bagger, {\em Supersymmetry and Supergravity},
2nd Edition, Princeton University Press, 1992.

\bibitem{kuz}
  I.~L.~Buchbinder and S.~M.~Kuzenko,
  {\em Ideas and methods of supersymmetry and supergravity: Or a walk through
  superspace},
Revised edition, IOP Publishing Ltd, 1998.





\end{thebibliography}
\end{document}